\documentclass[
    ,final            
  ]
  {aipproc}

\layoutstyle{6x9}

\usepackage{amsmath,amssymb}

\begin{document}

\title{Mirage unification at TeV scale and natural electroweak symmetry
breaking in minimal supersymmetry}


\author{Ken-ichi Okumura}{
  address={Department of Physics, Kyushu University, Fukuoka 812-8581, Japan}
}

\begin{abstract}
We propose a new minimal supersymmetric scenario
 with virtually no fine-tuning in the electroweak symmetry breaking.
It favors light supersymmetric spectrum below a few TeV
 and predicts definite relations among stop and gaugino
 masses and LSP higgsino with $\mu \lesssim 200$ GeV,
 which are within the reach of coming LHC experiment
\footnote{This talk is based on the collaboration in \cite{tevmirage}}.
\end{abstract}

\maketitle



Supersymmetry (SUSY) is a leading candidate for physics beyond the
standard model (SM) which solves the enigmatic gauge hierarchy problem.
The minimal extension leads to various non-trivial predictions: 
gauge coupling unification,
radiative electroweak (EW) symmetry breaking, 
the lightest SUSY particle (LSP) protected by R-parity
 (a cold dark matter candidate) and 
a light Higgs boson compatible with the EW precision tests. 
All of them are direct consequences of a single spell, 'minimal supersymmetry'.
This is a remarkable success, considering various difficulties we often encounter in alternative approaches. 

However, the LEPII experiment observed neither a Higgs boson
 nor a SUSY particle and has built up an uncomfortable tension
 in the EW symmetry
 breaking of the minimal supersymmetric standard model (MSSM).
At tree level, the lightest Higgs boson in the MSSM is lighter than $M_Z$.
Thus, the LEPII SM Higgs mass bound $m^2_{h^0} > 114.4$ GeV requires
 a substantial radiative correction, which raises the upper bound
 approximately, 
\begin{equation}
m^2_{h_0} \lesssim M_Z^2 + \left(3 g^2 m_t^4/8\pi^2 M_W^2 \right) \ln\left(m^2_{\tilde{t}}/m_t^2\right),
\end{equation}
 neglecting the trilinear scalar coupling, $A_t$. 
Then the SM Higgs mass bound is translated into $m_{\tilde{t}}\gtrsim 500$ GeV.
This bound for the stop mass is closely related to  
 that of the up-type Higgs, $H_u$ through the renormalization group (RG)
 running.
If the boundary value of $m^2_{H_u}$ is given at the cut-off scale $\Lambda$, 
a dominant correction below it
 is estimated as,
\begin{equation}
\Delta m^2_{H_u} \sim -\left(3/4\pi^2\right) y_t^2 m^2_{\tilde{t}} \ln\left(\Lambda/m_{\tilde{t}}\right).
\end{equation}
Then $\Lambda \approx M_{GUT}$ leads to $\Delta m^2_{H_u} \sim -2 m^2_{\tilde{t}}$, which suggests $m^2_{H_u} \sim {\cal
O}\left(m^2_{\tilde{t}}\right)$ up to a 'conspiracy' between
 $m^2_{H_u}(\Lambda)$ and the RG running.
While the EW symmetry breaking requires,
\begin{equation}
M_Z^2/2 = \left(m^2_{H_d}-m^2_{H_u}\tan^2\beta\right)/\left(\tan^2\beta-1\right)
 -|\mu|^2
\approx -m^2_{H_u} -|\mu|^2.
\end{equation}
Unless $m^2_{H_u} \sim \mu^2 \sim M_Z^2$, we need a fine-tuning between
 $m^2_{H_u}$ and $\mu^2$, despite their origins are totally different from each other.  The
 degree of fine-tuning is measured by,
\begin{equation}
\Delta({\mu^2})^{-1} \equiv \left(\partial
			    \ln M_Z^2/\partial \ln \mu^2\right)^{-1}
                    = M_Z^2/2|\mu|^2.
\label{ft}
\end{equation}
This implies $m_{H_u} \sim m_{\tilde{t}} \gtrsim 500$ GeV results in a fine-tuning worse than a few \% level.

Various solutions have been proposed for this 'little SUSY hierarchy
problem' between the EW scale and the SUSY (stop) scale.
They are classified in two categories.
The first solution raises the tree level Higgs mass well above $M_Z$ by
 enhancing the Higgs quartic coupling, which is solely determined
 by $SU(2)_L\times U(1)_Y$ D-term in the MSSM.
The second solution explores a way to realize
 the little hierarchy by introducing a new symmetry around the EW scale
 or lowering the cut-off $\Lambda$ close to the EW scale.
In both cases, new fields and thresholds are required.
This spoils the original simplicity of the minimal scenario, in particular, the prediction of the gauge coupling unification.

In this paper, we pursue an alternative direction, preparing the
conspiracy between $m^2_{H_u}$($M_{GUT}$) and the RG running below it,
in order to realize the EW scale little hierarchy \cite{tevmirage,tevmirage2}. This looks hard to achieve but
 is actually possible in the mirage mediation scenario of SUSY breaking
\cite{Choi:2004sx,Choi:2005uz,Endo:2005uy}, 
 typically realized in the KKLT-type string modulus
 stabilization\cite{Kachru:2003aw}. 
%
The system is described by the following effective supergravity action,
\begin{align}
\int d^4x & \left[\int d^4\theta (-3) CC^\ast \exp\left(-{\cal
 K}/3\right) +\left\{\int d^2\theta \left(\left(f_a/4\right) W^a W^a +
 C^3 {\cal W} \right) 
 + {\rm h.c.}\right\} \right. \nonumber\\ 
 & \left.
 ~+ \int d^4\theta C^2 C^{\ast 2} {\cal P}_{\rm lift} \theta^2
 \overline{\theta}^2 \right]
\end{align}
where $C= C_0 + F^C \theta^2$ denotes the chiral compensator superfield.
K\"ahler potential ${\cal K}$, gauge kinetic function $f_a$,
superpotential $\cal W$ and uplifting function ${\cal P}_{\rm lift}$ 
( represents the effect of SUSY breaking brane \cite{Kachru:2003aw,Choi:2004sx}) are
given by,
\begin{align}
& {\cal K} = {\cal K}_0(T+T^\ast) + Z_i(T+T^\ast) \Phi_i^\ast \Phi_i,~~~
 & f_a = T,~~~ \nonumber\\
& {\cal W} = w_0 - A e^{-a T} + \lambda_{ijk} \Phi_i \Phi_j \Phi_k,  & {\cal P}_{\rm lift} = {\cal P}_{\rm lift}(T+T^\ast),  
\end{align}
where $T$ denotes the gauge coupling modulus and $\Phi_i$ stands for chiral
matter superfields. 
In ${\cal P}_{\rm lift}=0$ limit, the modulus is stabilized at SUSY
AdS vacuum with $aT \approx \ln \left( A/w_0\right) \approx \ln \left(M_{Pl}/m_{3/2}\right)$, where 
$F^T \equiv -e^{{\cal K}/2} K^{T T^\ast} D_T^\ast {\cal W}^\ast = 0 $
 and the potential minimum is given by $\langle V \rangle = -3 |m_{3/2}|^2$.
This vacuum energy is fine-tuned to $\langle V \rangle \approx 0^+$ by
${\cal P}_{\rm lift}$. 
This uplifting process generates finite $M_0
\equiv F^T/(T+T^\ast)$ due to the shift of modulus vev \cite{Choi:2004sx}.
We introduce a parameter, $\alpha$ indicating
 the relative size of $m_{3/2}$ to $M_0$,
\begin{align}
\alpha \equiv m_{3/2}/M_0\ln(M_{Pl}/m_{3/2}) \approx
 \left\{1+\left(3/2\right)\left(\partial_T\ln {\cal P}_{\rm lift}/\partial_T {\cal K}_0\right)\right\}^{-1}.
\end{align}
This is typically ${\cal O}(1)$ and modulus mediated SUSY
breaking in the visible sector is comparable to the anomaly mediated one
\cite{Choi:2004sx}.
Note that the relative phase is dynamically eliminated \cite{susycp,Endo:2005uy}.
The soft SUSY breaking terms of the visible fields are summarized as,
\begin{align}
{\cal L} = -\left(1/2\right)
 M_a \overline{\lambda^a} \lambda^a - m^2_i |\phi_i|^2 - \left\{ \left(1/6\right)\, A_{ijk} y_{ijk} \phi_i \phi_j \phi_k  + {\rm h. c.}\right\}.  
\end{align}
where $\phi_i$ denotes a scalar component of $\Phi_i$, 
 and $\lambda^a$ represents gaugino. 
The canonical Yukawa coupling $y_{ijk}$ is defined as, 
 $y_{ijk} = \lambda_{ijk}/\sqrt{e^{-{\cal K}_0} Z_i Z_j Z_k}$.
These terms are derived at the unification scale using standard
 supergravity formula,
\begin{align}
& M_a = M_0 +\frac{b_a g_a^2}{16\pi^2} m_{3/2},~~~
 A_{ijk} = (a_i+a_j+a_k) M_0 -\frac{1}{16\pi^2}
\left(\gamma_i+\gamma_j+\gamma_k\right) m_{3/2}, \\
& m_i^2 = c_i M_0^2 -\frac{m_{3/2} M_0}{8\pi^2}
\left(2 C^a_a(\phi_i) g_a^2-\sum_{jk} \frac{1}{2}|y_{ijk}|^2 (a_i+a_j+a_k)\right)
-\frac{1}{32\pi^2}\frac{d \gamma_i}{d \ln\mu} |m_{3/2}|^2,
\nonumber\\
&\gamma_i = (8\pi^2)\frac{d \ln Z_i}{d \ln\mu},~~
a_i = 2 Re(T) \partial_T\ln\left(e^{-{\cal K}_0/3} Z_i \right),~~
c_i  = -4 Re(T)^2 \partial_T \partial_{T^\ast} 
 \ln\left(e^{-{\cal K}_0/3} Z_i \right), \nonumber        
\end{align}
where $a_i$ and $c_i$ are typically non-negative rational numbers.

Intriguing feature of this mirage mediation is that the anomaly mediated
contribution $(\propto m_{3/2},~ m_{3/2}^2 )$ effectively shifts
 the modulus mediation scale from $M_{GUT}$ to 
{\it the mirage messenger scale},
 $M_{\rm mir} \equiv M_{GUT}/(M_{Pl}/m_{3/2})^{\alpha/2}$ 
at one-loop level
\cite{Choi:2005uz}.
This holds for all the soft terms
 if the Yukawa coupling is negligible or only allowed for the
 combination of fields which satisfies $a_i+a_j+a_k=c_i+c_j+c_k=1$. 
It is straightforward to show
$M_a(\mu)=\left(g^2_a(\mu)/g^2_a(M_{\rm mir})\right)M_0 $ by explicit calculation.
While at $M_{GUT}$, $g_a^2 = g_G^2 = Re(T)^{-1}$ and $y_{ijk}^2 =\lambda^2 /e^{-{\cal K}_0}Z_i Z_j Z_k \propto
Re(T)^{-1}$, if the above condition is satisfied.
 Hence, the anomalous dimension scales like $\gamma_i\left(Re(T)\right)=Re(T)^{-1}\gamma_i(1)$ at
one-loop level.
This enables us to solve the $T$ dependence of $\ln Z_i$ at arbitrary scale as
 $\ln Z_i(Re(T))|_{\mu=M_{GUT}}+\Delta \ln Z_i((\mu/M_{GUT})^{1/g^2_G Re(T)})$.
Promoting $T$ as a superfield and replacing $M_{GUT}\to M_{GUT} \sqrt{CC^\ast}$,
it is straightforward to show $\Delta \ln Z_i|_{\theta^2}=\Delta
\ln Z_i|_{\theta^2 \overline{\theta}^2}=0$ at $M_{\rm mir}$, which leads to
$A_{ijk}(M_{\rm mir})=M_0$ and $m_i^2(M_{\rm mir})=c_i M_0^2$ in case of
 non-Abelian gauge group. Then the soft terms having the same
$c_i$ unify at $M_{\rm mir}$ ({\it mirage unification}).

This property is exactly what we need to realize the little hierarchy
between the Higgs and stop masses in the minimal SUSY.
Once we have $\alpha=2$, $a_{u_3}+a_{q_3}+a_{H_u}=1$
 and $c_{H_u}=0$, $c_{u_3}+c_{q_3}=1$
 due to discrete nature of the underlying physics,
we can realize $M_{\rm mir} \approx M_0$ and $m^2_{H_u}(M_{\rm mir}) \approx
 M_0^2/8\pi^2$ while
 $m^2_{\tilde{u}_3}(M_{\rm mir})+m^2_{\tilde{q}_3}(M_{\rm mir}) = M_0^2$.
Note that we do not add any new fields other than $T$ and the gauge
coupling constants unify at $M_{GUT}$.

An extra bonus of this scenario is natural realization of $\mu$ and $B$
 of order $M_0$, which
is non-trivial to achieve with the anomaly mediation due to
an unsuppressed ${\cal O}(m_{3/2})$ contribution to $B$. 
If we assume the $\mu$ term is generated by the same
non-perturbative dynamics as in the modulus stabilization, like
$\Delta {\cal W}=\tilde{A} e^{-aT} H_d H_u$, $\mu$ is naturally ${\cal
 O}(M_0)$ and the leading ${\cal O}(m_{3/2})$ terms in $B$ cancel
 at $\alpha=2$, leaving $B={\cal O}(M_0)$ \cite{Choi:2005uz,tevmirage}.
 
In this scenario, $H_d$ can belong to either the EW scale
($c_{H_d}=0$) or the SUSY scale ($c_{H_d}\neq 0$) as far as $\tan\beta$
 is not extremely large. In the former case, the EW symmetry
 breaking condition,
 $B\mu \approx (m^2_{H_d}+m^2_{H_u}+2\mu^2)/\tan\beta$ requires
$B \sim M_Z/\tan\beta$. This is beyond the control of the above
mechanism generating $B$, although it is not the problem of the effective
theory description itself. 
While, in the latter case, $B\sim M_0$ and virtually no fine-tuning is
required. It is known that this non-universal choice for $c_{H_d,H_u}$
  generates an effective $U(1)$ D-term contribution in the soft masses.
It is positive for $m^2_{H_u}$, however, sufficiently small to be driven
 negative by other corrections in similar size.
In the left panel of Fig.1, we show RG evolution of the soft masses
 for $c_{q,u,d,l,e} = 1/2$, $c_{H_d}=1$. Non-negligible bottom Yukawa
 coupling and the effective D-term contribution perturb
 the exact unification, however, a sufficient
 little hierarchy is realized at $\approx M_0$.
In the right panel of Fig.1, we plot the fine-tuning parameters
 $\Delta({\mu^2})^{-1}$ and $\Delta(|B|)^{-1}$ (similar definition
 as in (\ref{ft})) as a function of $m^2_{H_u}(M_0/\sqrt{2})$. Here
 $B_0$ denotes
 $B(M_{GUT})$ other than the loop induced anomaly mediated contribution.
Note that the EW symmetry is broken with
 fine-tuning better than  $10$ \% for predicted $m^2_{H_u}
 \sim M_0^2/8\pi^2$ with $M_0\approx M_{1,2,3} \gtrsim 1$ TeV while
the lightest Higgs mass  is comfortably accommodated above the SM bound.
In this scenario, LSP is always pure higgsino,
 typically lighter than $\sim$200 GeV. 








\begin{figure}
  \includegraphics[height=.23\textheight]{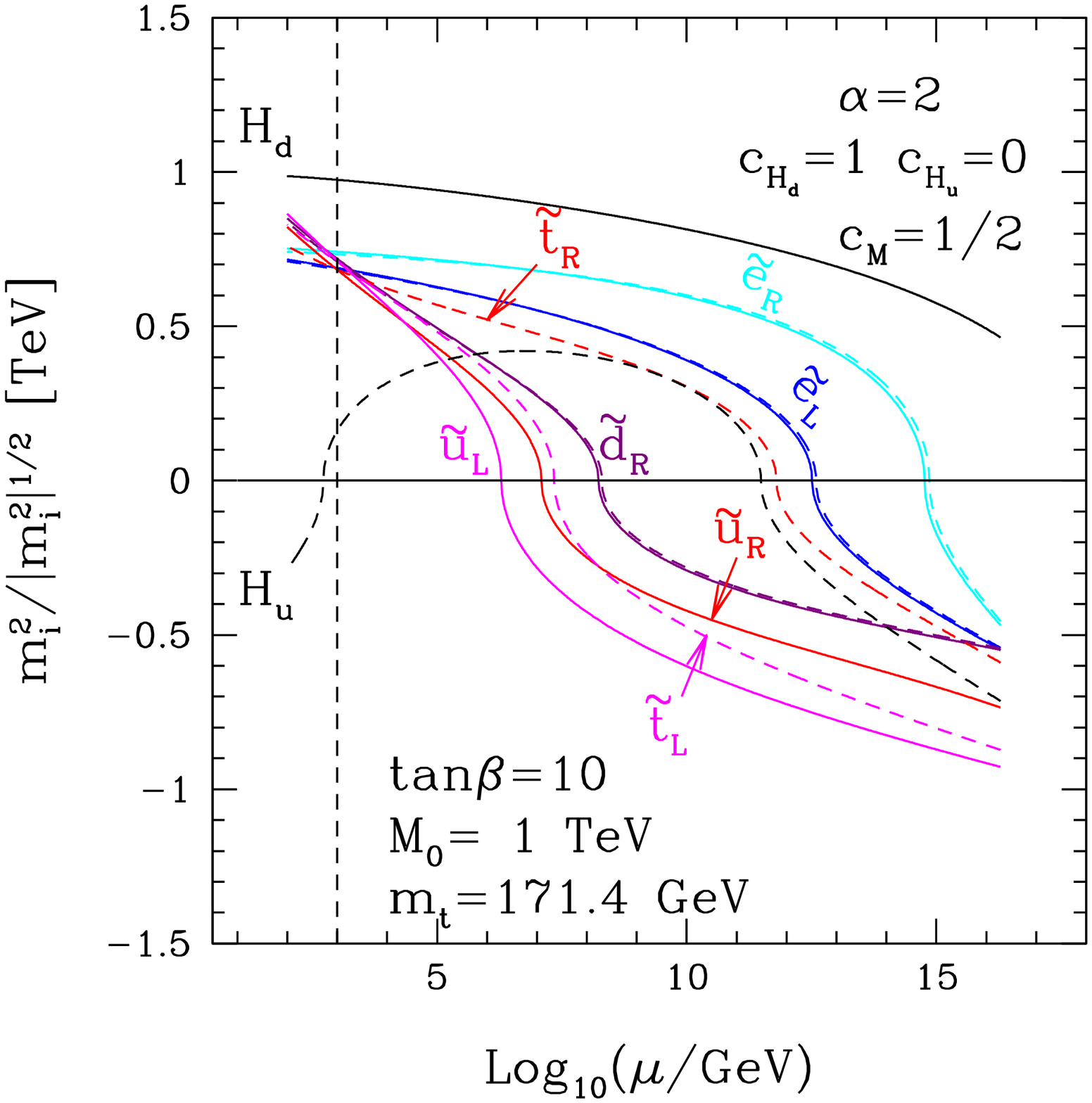}
  \includegraphics[height=.23\textheight]{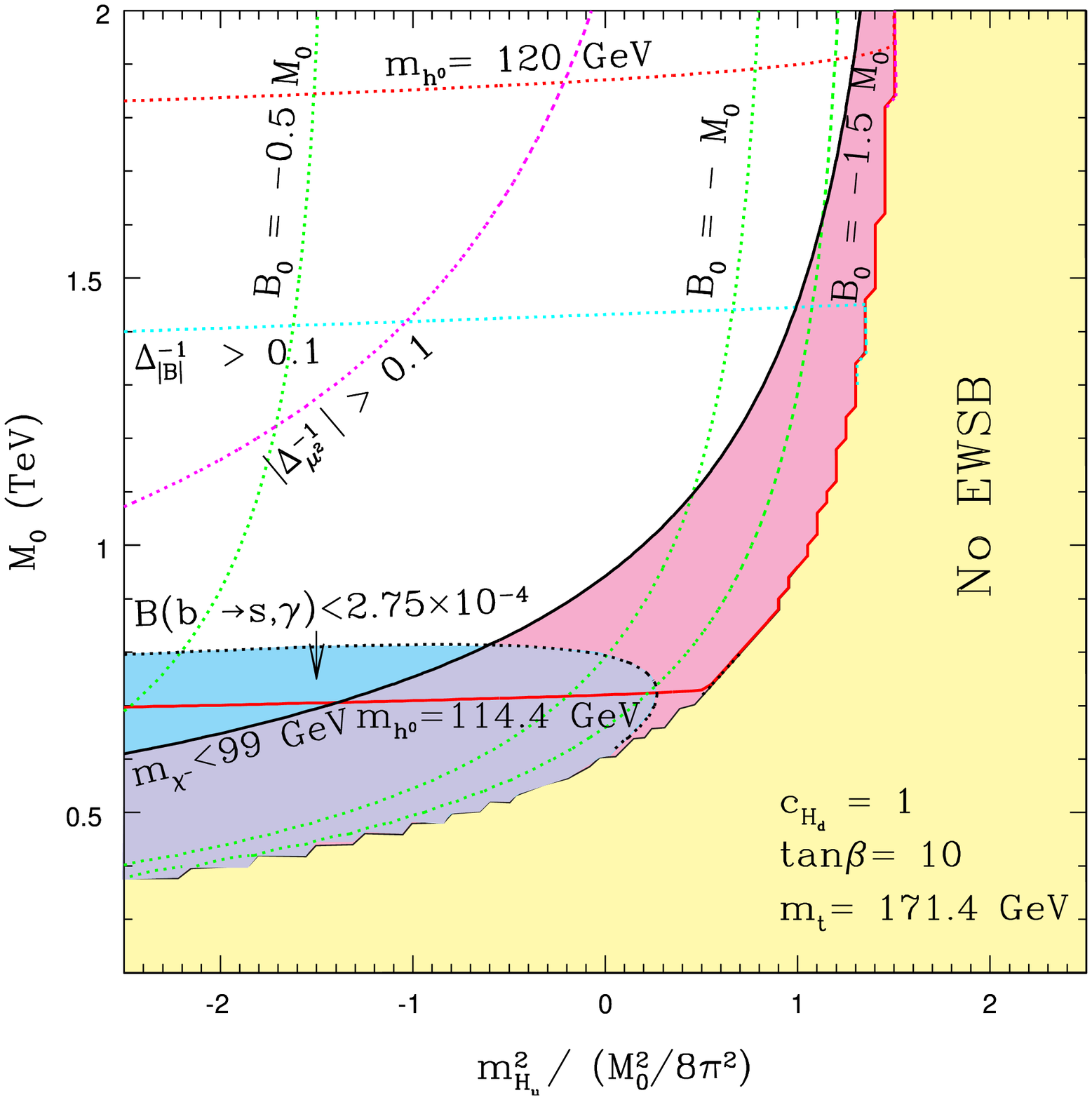}
  \caption{RG evolution and fine-tuning of the EW symmetry breaking in
 the TeV mirage mediation.}
\end{figure}

%


\begin{theacknowledgments}
This work is completed under the grant-in-aid for
scientific research on priority areas (No. 441): "Progress in
elementary particle physics of the 21 century through discoveries
of Higgs boson and supersymmetry" (No. 16081209) from 
the MECSST of Japan.
The numerical calculations were carried out on Altix3700 BX2
 at YITP in Kyoto University.
\end{theacknowledgments}




\bibliography{sample}


\end{document}